\documentstyle[12pt]{article}
\newcommand{\beeq}{\begin{equation}}
\newcommand{\eneq}{\end{equation}}
\newcommand{\beqn}{\begin{eqnarray}}
\newcommand{\eeqn}{\end{eqnarray}}
\def\la{\langle}
\def\ra{\rangle}

\begin{document}

\title{The General Correlation Function in the Schwinger Model on a Torus}

\author{S.Azakov \footnote{e-mail address: azakov\_s@hotmail.com}}
\maketitle

{\it Institute of Physics, Azerbaijan Academy of Sciences, Baku,
Azerbaijan}\\

\begin{abstract}
In the framework of the Euclidean path integral approach we derive the
exact formula for the general $N$-point chiral densities correlator
in the Schwinger model on a torus.
\end{abstract}
PACS number(s): 11.10 Mn \ldots Field theory, Schwinger source theory;\\
11.15 Tk \ldots Gauge field theories, other nonperturbative
techniques.\\
Keywords: Schwinger model, zeromodes, torus, fermionic correlator\\
Los Alamos Database Number: hep-th/0106071\\
Full postal address: Institute of Physics, Husein Javid ave.33,
Baku, 370143, Azerbaijan\\
Tel.99412 391424, fax 99412 395961

\section{Introduction}

The Schwinger model \cite{Schwin} (SM) (two-dimensional QED with
massless fermions) on a Euclidean torus ${\cal T}_2$ is exactly
soluble \cite{Azakov} \cite {JA}, \cite {SW} and in many
calculations it would be useful to have an expression for the
general $N$-point correlation function of chiral densities.

It is well known that in this model the "photon" acquires a mass due to chiral
anomaly and fermions disappear from the physical spectrum.

There are some features of the SM which have similiarity with those of QCD.
Fermion condensate, mass generation, dynamical symmetry breaking and
confinement are among them. In both models instantons are supposed to be
responsible for some nontrivial vacuum expectation values \cite{instQCD},
\cite{Smilga1}, \cite{SVZ}.

Work on a torus is desirable by several reasons.
Firstly, by defining the model on a finite volume we get rid of
infrared problems. Compactification makes mathematical manipulations more
rigorous, topological relations become more precise and transparent.
One should remember that the fermion path integrals have no meaning
unless defined using a discrete basis.
\par
Secondly, in this case we have a model with nontrivial topology in which we
can find explicitly fermionic zero modes and Green's functions in all
topological sectors. A presence  of topologically non-trivial configurations
of the gauge field (instantons) and fermionic zero modes allows in path
integral framework to reproduce the structure of the SM found in the operator
formalism \cite{LS}.
\par
Thirdly,  a compactification on a torus allows to find finite temperature
and finite size effects and is appropriate to the systematic analysis
of the lattice approximtion. It is a torus which is most naturally
approximated by the finite cubical lattice on which the numerical calculations
are performed \cite{latSM}.
\par
Finally, torus and circle are particularly appropriate for studying the
relation between the Hamiltonian and path integral approach
in the gauge theory with massless fermions \cite{JoosSP}, \cite{JAinprep}.

Thanks to its full solubility the SM may also be used to test various ideas
related to nonperturbative structure of quantum field theory.

Bardakci and Crescimanno were the first who using path
integral approach explored the role of nontrivial topological configurations
in two-dimensional fermionic model relevant in the context of string
compactification \cite{BC}.
They were able to show that certain correlation functions which,
being zero for the trivial topology, considerably change by nontrivial
topological effect. Following the ideas of Bardakci and Crescimanno
Manias, Naon and Trobo studied the behaviour of correlation functions of
fermion  bilinear operators in the SM with topologically nontrivial gauge
configurations in the infinite space-time \cite{Naon}.
Two- and four- point correlation functions in the SM on the torus have been
calculated in \cite{Azakov},\cite {FHNVZ}. The authors of the paper \cite{SSZ}
considered a six-point correlation function in this model which
due to some technical difficulties they managed to calculate only at finite
temperature but in the infinite space. They also made a conjecture about the
explicit expression for the N-point correlator of chiral densities again at
the finite temperatures but in the infinite space.

The paper is organized as follows. In Section 2 we briefly review the
results obtained before for the SM on the torus in Euclidean (path
integral) approach \cite {Azakov},\cite{JA},\cite{SW} and relevant for the
present consideration.
In addition to these results we give some new information
which concerns the possible choice of the zero modes and fermionic Green's
functions in the nontrivial topological sectors. Section 3 is devoted to the
discussion of modular transformation. The invariance with respect to this
transformation helps us in the sequel to determine some proportionality
constants. In section 4 which is the central part of the present work we
get our main result.
The last section is reserved for conclusions and the discussion
of possible directions of the future investigation.

\section{A brief review of path integral formulation of the SM on the
Euclidean torus}
The SM action on the Euclidean torus ${\cal T}_2, (0\leq
x_1\leq L_1, 0\leq x_2 \leq L_2)$ reads
\beeq
S = \int_{{\cal T}_2 }\,d^2x\left(\frac{1}{2}F_{12}^2(x) +
\bar{\psi}(x)\gamma_{\mu}(\partial_{\mu} - i\,e A_{\mu}(x))\psi(x)\right)~,
\label{action}
\eneq
where $F_{12}(x) = \partial_1A_2(x) - \partial_2A_1(x)$ is a field strength.
We use conventions and notations used in \cite {JA},\cite {Azakov}.
The evaluation of quantum mechanical expectation values (QMEV) in the
path integral formulation
\beeq
\langle \Omega [\bar{\psi},\psi, A_{\mu}]     \rangle =\frac{1}{Z}
 \int {\cal D}[\psi,\bar{\psi},A_{\mu}] \Omega [\bar{\psi},\psi, A_{\mu}]
 e^{-S[\bar{\psi},\psi,A_{\mu}]}~,
\label{qmev}
\eneq
where the $Z$ factor (the partition function):
\beeq
Z=\int {\cal D}[\psi,\bar{\psi}, A_{\mu}]e^{-S[\bar{\psi},\psi, A_{\mu}]}~,
\label{Zfactor}
\eneq
gets the following form for QMEV of fermion fields \cite{Azakov}
\beqn
&&\langle\psi_{\alpha_1}(x_1)\bar{\psi}_{\beta_1}(y_1)\cdots\psi_{\alpha_N}
(x_N)\bar{\psi}_{\beta_N}(y_N)\rangle \nonumber \\
&&=Z^{-1}\sum_{k=0,\pm1,\cdots
,\pm N}L_1^{|k|}\int_{{\cal A}_k}{\cal D}Ae^{-S[A]} {\det}'[L_1 \gamma_{\mu}
(\partial_{\mu} -ieA_{\mu})] \nonumber \\
&&\times \sum_{P_i}(-1)^{p_i}\sum_{P_j}(-1)^{p_j}
\hat{\chi}_{\alpha_{i_1}}^{(1)}(x_{i_1})\cdots
\hat{\chi}_{\alpha_{i_{|k|}}}^{(|k|)}
(x_{i_{|k|}})\bar{\hat{\chi}}_{\beta_{j_1}}^{(1)}(y_{j_1})\cdots
\bar{\hat{\chi}}_{\beta_{j_{|k|}}}^{(|k|)}(y_{j_{|k|}}) \nonumber \\
&&\times
S_{\alpha_{i_{|k|+1}}\beta_{j_{|k|+1}}}^{(k)} (x_{i_{|k|+1}},
y_{j_{|k|+1}};A)
\cdots
S_{\alpha_{i_N} \beta_{j_N}}^{(k)}(x_{i_N},y_{j_N};A)~.
\label{gfor}
\eeqn
Here we have already performed the fermion integration
$\int {\cal D}[\psi,\bar{\psi}]$ over the fermionic \\
Grassmann variables. The sum is taken over all possible permutations\\
$P_i =(i_1,i_2,\ldots,i_N)$ and $P_j =(j_1,j_2,\ldots,j_N)$ of
$(1,2,\ldots,N)$ ,  $(-1)^{p_i} ((-1)^{p_j})$
is a parity of the permutation $P_i (P_j)$.
\par
Let us discuss the topological origin and the physical meaning of
the different parts of this expression for  the SM on the 2d torus:
\subsection {Topology of $U(1)$-gauge fields $A_{\mu}(x)$ on the torus }
The topology of $U(1)$-gauge fields on ${\cal T}_2$ :
\beeq
A_{\mu}(x) = C^{(k)}_{\mu}(x) + t_{\mu} + \epsilon_{\mu \nu}
\partial_{\nu}b(x) +\partial_{\mu} a(x)
\label{Hodge}
\eneq
is given by the decomposition of $A_{\mu}$ into Chern classes
together with the Hodge decomposition \cite{index}.
$C^{(k)}_{\mu}(x) = -\frac{2 \pi k}{e L_1 L_2} x_2 \delta_{\mu,\nu}$,
a gauge potential in the Lorentz gauge which leads to a constant field strength
$F_{\mu\nu}(x) = \frac{2\pi k}{eL_1L_2} \epsilon_{\mu\nu}$
of the stationary gauge action. It belongs to the Chern class
with a topological charge (topological quantum number)
$\frac{e}{2\pi}\int_{{\cal T}_2}F_{12}d^2x =k $,
and it plays the role of an {\bf instanton} in our model. It defines a
connection of a principal non-trivial $U(1)$-bundle over  ${\cal T}_2$ with transition
functions $\Lambda_{\nu}(x)$:
\[A_{\mu}(x + \hat{L}_{\nu})= A_{\mu}(x) - \frac{i}{e}\Lambda^{-1}_{\nu}
(x)\partial_{\mu}\Lambda_{\nu}(x)~.\]
In our gauge the transition functions are gauge transformations:
\beqn
\Lambda_1(x) = e^{\pi i k \frac{x_2}{L_2}},\;\;\;\;\;
\Lambda_2(x)= e^{-\pi i k \frac{x_1}{L_1}}~,
\label{tranfun}
\eeqn
and describe the continuation of $C^{(k)}$ in the $U(1)$ bundles along a cycle
in ${\cal T}_2$. $t_{\mu} $ is a harmonic potential: $\Box t_{\mu} = 0 $,
called toron field. It is a zero mode of the gauge field
and is restricted to $0\leq t_{\mu}\leq T_{\mu}$,
where $T_{\mu}\equiv \frac{2\pi}{eL_{\mu}}$.
$\epsilon_{\mu \nu}\partial_{\nu} b(x)$ describes
gauge independent `deformations ' of $C^{(k)}_{\mu}(x)$,~~$a(x)$ is a pure
local gauge: $ \partial_{\mu}a(x) =-\frac{i}{e}
e^{ -iea(x)}\partial_{\mu} e^{ iea(x)}$.{\bf Large gauge transformations} on
${\cal T}_2$: $\Lambda(x)
= \exp2\pi i(m_1\frac{x_1}{L_1}  + m_2 \frac{x_2}{L_2})\; $
transform the toron field according to $t_{\mu} \rightarrow
t_{\mu} +T_{\mu}m_{\mu}$.
The Hodge decomposition leads to a product decomposition of the
functional measure appearing in the path integral formulation:
\beqn
\int {\cal D}A = \sum_k \int {\cal D}A^{(k)}=
\sum_k \int^{T_{\mu}}_0 dt_{\mu}\int {\cal D}a \int {\cal D}b
\label{funcm}
\eeqn
\subsection{Fermionic zero modes $\chi(x)$}
The Atiyah-Singer index theorem for Dirac equation
\cite{index} states that the  number
of solutions with spin parallel minus the number of solutions
with spin anti-paralell  is equal to $|k|$.
The fermionic zero modes are solutions of the Dirac equation
satisfying the periodic boundary conditions described by the
transition functions $\Lambda_{\nu}(x)$ of the $U(1)$-bundle:
\beqn
\label{dirac1}
\gamma_{\mu}(\partial_{\mu} -ieA_{\mu})\hat{\chi}(x) =0
\;\;\; {\rm with} \;\; \hat{\chi}(x+ \hat{L}_{\nu}) = \Lambda_{\nu}(x)
\hat{\chi}(x).
\eeqn
In future we will consider also the operator
\beqn
D_0=D\vert_{a=b=0}=\gamma_{\mu}(\partial_{\mu}-ie(t_{\mu}+C_{\mu}^{(k)})))~.
\eeqn
Then $\hat{\chi}^{(j)}(x) = e^{iea(x)+e\gamma_5b(x)}\chi^{(j)}(x)$, where
$\chi ^{(j)}(x)$ is a zero mode of the $D_0$ operator, which can be
explicitly expressed by Jacobi's $\theta$-functions \cite{theta}, \cite{Bat}.
 The  most general expression for the zero modes of the $D_0$
operator with the positive chirality $(k>0)$ in the Lorentz gauge
takes a form \cite{Joos} $(j=1,\ldots,k)$ \beqn \chi ^{(j)}(x) =
\left(\matrix{ \chi_1^{(j)}(x) \cr 0\cr}\right) \eeqn with \beqn
\chi_1^{(j)} (x) =  \left(\frac{2k}{|\tau|}\right)^{1/4}
\frac{1}{L_1}e^{\frac{2\pi i}{|\tau|}\zeta\bar{t} +\frac{i\pi
k}{|\tau|}z\zeta -\frac{i\pi}{k}\bar{t}\tilde{t}_1}T_k^{(j)}(z'),
\label{zeromode} \eeqn where functions $T^{(j)}_k(z)$ obey the
periodicity conditions \beqn T^{(j)}_k(z+1)=T^{(j)}_k(z)~, ~~~~
T^{(j)}_k(z+\tau)=e^{-i\pi k(2z+\tau)}T^{(j)}_k(z)
\label{Tpercond} \eeqn and are chosen in such a way that the zero
modes Eq.(\ref{zeromode}) are orthonormalized ($z' \equiv z
+\bar{t}/k$, where $z\equiv\frac{x_1+ix_2}{L_1}, \zeta \equiv{\rm
Im}z, \tau \equiv i \frac{L_2}{L_1}$ and $t\equiv\tilde{t}_2
+i|\tau|\tilde{t}_1$,
$\tilde{t}_{\mu}\equiv\frac{eL_{\mu}}{2\pi}t_{\mu}$, bar means
complex conjugation). The constant factor in the r.h.s. of
Eq.(\ref{zeromode}) is chosen for convenience. We have two
explicit solutions. One was presented in \cite{Azakov} \beqn
T_k^{(j)}(z')=e^{-\frac{\pi(j-1)^2}{k}|\tau|+2\pi
i(j-1)z'}\theta_3(kz' +(j-1)\tau|k\tau). \label{set1} \eeqn
Another has a form: \beqn
T_k^{(j)}(z')=\frac{1}{\sqrt{k}}\theta_3\left(z'-\frac{(j-1)}{k}\left\vert
\frac{\tau}{k}\right.\right)~. \label{set2} \eeqn Note that this
is actually the solution found by Sachs and Wipf \cite{SW}. In
order to see this one should apply to it the modular
transformation considered in Section 3.

Of course, each function of the set Eq.(\ref{set2}) is a linear combination
of the functions of the set Eq.(\ref{set1}), since there is a relation
\beqn
\theta_3(z|\tau/k)=\sum_{l=0}^{k-1}e^{\pi i l^2\frac{\tau}{k} +2\pi i lz}
\theta_3(kz+l\tau|k\tau) ~.
\eeqn
For the zero modes of the negative chirality ($k<0$) we have
\beqn
\phi ^{(j)}(x) = \left(\matrix{
0\cr
\phi_2^{(j)}(x) \cr}\right)~,~~~~j=1,\ldots,|k|
\eeqn
with
\beqn
\phi_2^{(j)} (x) =  \left(\frac{2|k|}{|\tau|}\right)^{1/4}
\frac{1}{L_1}e^{-\frac{2\pi i}{|\tau|}\zeta t
-\frac{i\pi |k|}{|\tau|}\bar{z}\zeta
+\frac{i\pi}{|k|}t\tilde{t}_1}T_{|k|}^{(j)}(\bar{z}^{\prime\prime}),
\label{zeromode1}
\eeqn
where $\bar{z}^{\prime\prime}\equiv\bar{z} -\frac{t}{|k|}$.
\subsection{Regularized effective  action
$\Gamma^{(k)}_{reg}[A]$}

 We have calculated  the regularized effective action \\
$\Gamma^{(k)}_{reg}[A] = 2\ln \det^{\prime}(L_1 \gamma_{\mu}
(\partial_{\mu} - ieA_{\mu}))+ \Gamma_{reg}(\{M_j\})$ for the different
topological sectors. The result is \cite{Azakov}, \cite{JA}:
\begin{eqnarray}
\Gamma_{reg}^{(k)}[A] &=& \frac{e^2}{\pi} \int_{{\cal T}_2} d^2x b(x)
\Box b(x)\nonumber\\
& + &  2\delta_{0,k}\ln |e^{-\frac{2\pi}{|\tau|}\tilde{t}^2_1}
\theta_1(t|\tau)\eta^{-1}(\tau)|^2
\label{Gamma}\\
&-& (1-\delta_{0,k}) \Big \{ 2\ln \det{\cal N}_A^{(k)}
- \vert k \vert (\ln(2 \vert k \vert / |\tau|-2\pi i)) \Big\} \nonumber \\
&  + &\Gamma_{reg}(\{M_j\}) ~.\nonumber
\end{eqnarray}
As discussed below the first term is a `mass term',
see Eq.(\ref{scalprop}).
The second term defines the `induced toron action'
${\Gamma}^{(0)}[t]$ , ($\eta(\tau)$ is
Dedekind's function). It is induced by the fermions via the spectral
flow of the Dirac operator \cite{JAinprep}.
In calculating the effective action for gauge fields
from the topological non-trivial sectors $k \neq 0 $, one
has to separate the zero modes. The third line contains the
determinant of the matrix of the scalar products of the
(non-orthonormal) zero-modes $ {\cal N}_A^{(k)}$, and
a weight factor of the non-trivial sectors \cite{JA},\cite{Azakov}.
The regularization term $\Gamma_{reg}\{(M_j\})$
drops off by the normalization of the path integral formula.
The term $|k|(\ln(2|k|/|\tau| -2\pi i)$ compensates the length scale
dependence of the zero mode's normalization. It determines the relative
weights of the contributions from the different topological sectors.
Observe that in the general formula Eq.(\ref{gfor}) it is assumed that zero
modes ${\hat \chi}^{(j)}(x)$ are orthonormalized. If not the matrix
${\cal N}^{(k)}_A$ will enter this formula (see Eq.(\ref{rs1})).
\subsection{ The fermion propagator $ S^{(k)}(x,y;A)$}
It follows from the well-known solution of the 2d Dirac equation with
external gauge potential that the fermion propagator can be written as
\beeq
\label{fermprop}
 S^{(k)}(x,y;A) = e^{ie\alpha(x)} S_t^{(k)}(x,y)e^{-ie\alpha^{\dag}(y)}~,
\eneq
with $\alpha(x)=a(x)-i\gamma_5 b(x)$, where $S_t^{(k)}(x,y)$ is a
propagator of fermions in the background  gauge field $A_{\mu}(x),a=b=0$~ from
the sector with the topological charge $k$. There is an explicit expression
for $S^{(0)}_t(x,y)$ in terms of $\theta$-functions
\footnote
{In what follows we will use shorthand notations for
$\theta$-functions $\theta_{\alpha}(z)\equiv \theta_{\alpha}(z|\tau)$
if the second argument of a theta function is $\tau$.}
 \cite{Azakov}, \cite{JA}, \cite{FHNVZ}:
\beqn
S_t^{(0)}(x-y) =
\left( \begin{array}{cc}
 0 & \frac{\eta^3}{L_1}\frac{\theta_1(z-w+ \bar{t})}
 {\theta_1(\bar{t})\theta_1(z-w)} e^{\frac{2\pi i}{|\tau|}(\zeta -\xi)
\bar{t}}\\
 -\frac{\eta^3}{L_1}\frac{\theta_1(\bar{z}-\bar{w} - t)}
 {\theta_1(t)\theta_1(\bar{z}-\bar{w})} e^{\frac{-2\pi i}{|\tau|}
(\zeta-\xi)t}& 0
 \end{array} \right)~,
\label{fpt0}
\eeqn
where $w\equiv\frac{y_1+iy_2}{L_1}$ and $ \xi \equiv{\rm Im} w$.
Note that $ S_t^{(0)}(x)$ becomes singular for $t = 0$. This singularity
is caused by the constant solution of the Dirac equation with $t=0$.
It represents a zero mode in the trivial sector. In the path
integral it is compensated by a zero of
the Boltzmann factor of the induced
toron action: $ \sim \exp (\Gamma^{(0)}[t]/2) $.
In the sector with $k>0$ the fermionic Green's function takes a form
\cite{Azakov},\cite{SSZ}:
\beqn
S_t^{(k)}(x,y)=S_t^{(0)}(x,y)\frac{q^{(k)}(z)}{q^{(k)}(w)}e^{\frac{i\pi k}
{|\tau|}(z\zeta -w\xi)}~,
\label{fptk}
\eeqn
where $q^{(k)}(z)$ is a function which obeys the same periodicity
conditions as the functions $T_k^{(j)}(z)$ (see Eq.(\ref{Tpercond}))
\beqn
q^{(k)}(z+1)=q^{(k)}(z)~,~~~q^{(k)}(z+\tau)= e^{-i\pi k(2z+\tau)}q^{(k)}(z)~,
\label{qpercond}
\eeqn
and have no poles in $z$. Some examples are as follows:
\beqn
q^{(k)}(z)=\theta_3(z|\tau/k)
\label{qkz}~~,
\eeqn
\beqn
q^{(k)}_l(z)=e^{2\pi ilz}\theta_3(kz+l\tau|k\tau)~,
\label{qlkz}
\eeqn
where $l=0,1,\ldots,k-1$. Note  that any linear combination of functions
Eq.(\ref{qlkz}) also obeys the periodicity conditions Eq.(\ref{qpercond}).
When $k \neq 0$ the choice of the fermionic Green's function is not unique.
All possible choices differ by the linear combination of the zero modes.
\subsection{ Scalar propagators on the torus}
The $b(x)$- dependent part of the action consists of the gauge field action
$S_g[A]$ and the mass term of $\Gamma_{reg}^{(k)}[A]$ giving together
$S[b] = 1/2 \int_{{\cal T}_2}dx b(x)\Box(\Box - m^2)b(x) $ with
$m^2 \equiv e^2/\pi$. The corresponding propagator satisfies the equation:
\begin{equation}
\Box(\Box -m^2)G(x-y)
 = \delta^{(2)}(x-y) -\frac{1}{L_1L_2}~,
\label{scalprop}
\end{equation}
where $\delta^{(2)}(x-y)$ is Dirac's $\delta$-function on the torus.
It can be written as the difference of a massless and massive propagator
on the torus orthogonal to the constant functions:
$   G(x)=  1/m^2\{ G_0(x) - G_m(x)\}$. There is a closed expression in
the massless case
\beqn
G_0(x)= -\frac{1}{2\pi} \ln \left( \eta^{-1}(\tau)
e^{-\frac{\pi \zeta^2}{|\tau|}}|\theta_1(z)|\right)~.
\label{G0prop}
\eeqn
In the massive case we use the infinite sum for
$\overline{G}_m(x)= G_m(x) + 1/m^2 L_1 L_2 $:
\beqn
\overline{G}_m(x)=\frac{1}{2 L_1 }\sum_n
\frac{\cosh[ E(n)(L_2/2 - | x_2 |)]
e^{2\pi in \frac{x_1}{L_1}}} {E(n)\sinh [L_2 E (n)/2]},
\label{mprop}
\eeqn
where
$$E(n) = \sqrt{4 \pi^2n^2 L_1^{-2} + m^2}~.$$

\subsection{Chiral condensate and two-point correlators of chiral densities}
If one considers QMEV only of gauge invariant quantities the pure gauge
field $a(x)$ may be integrated over with no consequence and we will not
consider it in future. Then in the topological sector with the topological
charge $k$ we may write:
\beqn
\int_{{\cal A}_k}{\cal D}Ae^{-S[A]}\ldots=
e^{-\frac{2\pi k^2}{m^2L_1L_2}}\int_0^{T_1}dt_1\int_0^{T_2}dt_2
\int{\cal D}b e^{-\frac{1}{2}\int b(x)\Box^2b(x)d^2x}\ldots
\label{integ}
\eeqn
The partition function Eq.(\ref{Zfactor}) is a product of three factors:
\beqn
Z=\int_{{\cal A}_0}{\cal D}Ae^{-S[A]+\frac{1}{2}\Gamma_{reg}^{(0)}[A]}=
Z_0Z_tZ_M~,
\label{partfunc}
\eeqn
where
\beqn
Z_0 =\int{\cal D}be^{-\frac{1}{2}\int d^2xb(x)\Box(\Box-m^2)b(x)}~,
\label{pfb}
\eeqn
\beqn
Z_t=\int_0^{T_1}dt_1\int_0^{T_2}dt_2e^{\frac{1}{2}\Gamma^{(0)}(t)}
=\frac{(2\pi )^2}{e^2\sqrt{2|\tau|}L_1L_2\eta^2(\tau)}~,
\label{pft}
\eeqn
\beqn
Z_M=e^{\frac{1}{2}\Gamma_{reg}(\{M_j\})}~.
\label{pfM}
\eeqn
The QMEV of fermion fields exhibit the mechanism of chiral
symmetry breaking by an anomaly.
The configurations with $k=\pm 1$ are responsible for the formation of the
chiral condensate
\cite {JA}, \cite{SW}, \cite{Azakov}, which has a form:
\beqn
\langle \bar{\psi}(x)P_{\pm}\psi (x)\rangle =
-\frac{\eta^2(\tau)}{L_1}e^{2e^2 G(0) -\frac{2\pi ^2}{e^2L_1L_2}}~,
\label{fc}
\eeqn
where $P_{\pm}\equiv \frac{1}{2}(1\pm \gamma_{5})$,
and two-point correlators of chiral densities
\beqn
\langle\bar{\psi}(x)P_{+}\psi (x)\bar{\psi}(y)P_{-}\psi (y)\rangle
=(\langle \bar{\psi}(x)P_{+}\psi (x)\rangle)^2e^{4\pi \overline{G}_m(x-y)}~,
\label{chdencor1}
\eeqn
\beqn
\langle\bar{\psi}(x)P_{+}\psi (x)\bar{\psi}(y)P_{+}\psi (y)\rangle
=(\langle \bar{\psi}(x)P_{+}\psi (x)\rangle)^2e^{-4\pi \overline{G}_m(x-y)}
\label{chdencor2}
\eeqn
get contributions from the topological sectors with $k=0$ and $k=2$,
respectively \cite{Azakov}, \cite{FHNVZ}.
The expressions for this correlators of the SM in the infinite space-time
were first obtained by Casher, Kogut and Susskind \cite{CKS}
who used the bosonization techniques in the operator formalism. The nonvanishing
$\la {\bar \psi}(x)\psi(x) \ra^2 $ in this model was obtained for the first
time by Lowenstein and Swieca \cite {LS}.
\section{Modular transformation}
Under exchange
\beqn
(x_i)_1\leftrightarrow (x_i)_2,~~
L_1\leftrightarrow L_2,~~ t_1 \leftrightarrow t_2,~~ \gamma_1\leftrightarrow
\gamma_2
\eeqn
we have a modular transformation:
\beqn
\tau \rightarrow -\frac{1}{\tau},~~
z \rightarrow -\frac{\bar{z}}{\tau},~~
\bar{z} \rightarrow \frac{z}{\tau}~~,\cr
t \rightarrow -\frac{\bar{t}}{\tau},~~
\zeta \rightarrow \frac{\bar{z}+z}{2|\tau|},~~
\gamma_5 \rightarrow -\gamma_5~.
\label{modtran}
\eeqn
Under this transformation we have transitions
\beqn
&&\eta(\tau)\rightarrow \eta(-\frac{1}{\tau})=
\sqrt{|\tau|}\eta(\tau)~~,\cr
&&\theta_1(z|\tau)\rightarrow \theta_1(-\frac{\bar{z}}{\tau}|-\frac{1}{\tau})=
i\sqrt{|\tau|}e^{\frac {i\pi \bar{z}^2}{\tau}}\theta_1(\bar{z}|\tau)~,\cr
&&\theta_1(z+\bar{t}|\tau)\rightarrow \theta_1\left(-\frac{\bar{z}}{\tau}
+\frac{t}{\tau}|-\frac{1}{\tau}\right)=
\sqrt{|\tau|}e^{\frac {\pi \bar{z}^2}{|\tau|}-\frac{2\pi \bar{z} t}{|\tau|}+
\frac{\pi t^2}{|\tau|}}\theta_1(\bar{z}-t|\tau).
\eeqn
One can easily check that under modular transformation:
$G_0(x)$ and $S_t^{(0)}(x)$ are invariant, there is an exchange of
the zero modes of opposite chirality: $\chi^{(j)}_1(x) \leftrightarrow
\phi^{(j)}_2(x)$ and
\beqn
e^{\frac{i\pi k}{|\tau|}z\zeta}\theta_3(z|\tau/k)
\rightarrow \sqrt{|\tau|k}e^{-\frac{i\pi k}{|\tau|}\bar{z}\zeta}
\theta_3(k\bar{z}|k\tau)~,
\eeqn
\beqn
e^{\frac{i\pi k}{|\tau|}z\zeta}\theta_3(kz|k\tau)
\rightarrow \sqrt{\frac{|\tau|}{k}}e^{-\frac{i\pi k}{|\tau|}\bar{z}\zeta}
\theta_3(\bar{z}|\tau/k)~.
\eeqn
So the modular transformation acts on the fermionic Green's function in
the nontrivial topological sector Eq.(\ref{fptk}) effectively (up to a linear
combination of zero modes) as a complex conjugation.
\section{General formula}
The general formula which we want to prove is
\beqn
\la \prod_{i=1}^N\bar{\psi}(x_i)P_{e_i}\psi(x_i)\ra
=\left(\la\bar{\psi}(x)P_+\psi(x)\ra \right)^Ne^{-4\pi\sum_{i<j}
e_ie_j\overline{G}_m(x_i-x_j)}~,
\label{genfor}
\eeqn
where $e_i=\pm (\pm 1)$.
\par
Without loss of the generality we may consider two cases
($N=r+s, r-s=k, s\leq r$, $r(s)$ is a number of factors in the l.h.s. of
Eq.(\ref{genfor}) with $e =+ (e =-)$ :\\
1) $r=s$. Only a trivial sector $k=0$ contributes. From the most
general formula Eq.(\ref{gfor}) it follows that
\beqn
&&\la \prod_{i=1}^r\bar{\psi}(x_i)P_+\psi(x_i)\bar{\psi}(y_i)P_-\psi(y_i)\ra\cr
&&=Z^{-1}\int_{{\cal A}_0}{\cal D}Ae^{-S[A]+
\frac{1}{2}\Gamma^{(0)}_{\rm reg}[A]}
\left\vert \det
\left \Arrowvert S_{12}^{(0)}(x_i,y_i;A)\right\Arrowvert
\right\vert^2~.
\label{k01}
\eeqn
(We omit the matrix indices 12 of the $2\times 2$ fermion propagator matrix
in the following for shorthand. Since we consider only the case with $k\geq 0$
only this matrix element will be appearing in our calculations).
With the help of the Eq.(\ref{fermprop}) we may write
\beqn
\left\vert \det\left\Arrowvert S(x_i,y_j;A) \right\Arrowvert\right
\vert^2=e^{2e\sum_{i=1}^r[b(x_i)-b(y_i)]}
\left\vert \det\left\Arrowvert S_t^{(0)}(x_i,y_j)\right\Arrowvert\right
\vert^2~,
\label{detSA}
\eeqn
and do the path integration with over the $b$ field with the result:
\beqn
&&\la \prod_{i=1}^r\bar{\psi}(x_i)P_+\psi(x_i)\bar{\psi}(y_i)P_-\psi(y_i)\ra
=Z^{-1}_te^{N\left(2e^2G(0)-\frac{2\pi ^2}{e^2L_1L_2} \right)}\cr
&&\times e^{-4\pi\sum_{i<i'}^r
\overline{G}_m(x_i-x_{i'})-4\pi\sum_{j<j'}^r\overline{G}_m(y_j-y_{j'})
+4\pi\sum_{i=1}^r\sum_{j=1}^r\overline{G}_m(x_i-y_j)}\cr
&&\times e^{4\pi\sum_{i<i'}^r
G_0(x_i-x_{i'})+4\pi\sum_{j<j'}^r G_0(y_j-y_{j'})
-4\pi\sum_{i=1}^r\sum_{j=1}^r G_0(x_i-y_j)}\cr
&&\times \int_0^{T_1}dt_1\int_0^{T_2}dt_2
\left |\det \left \| S^{(0)}_t(x_i,y_j) \right \|
\right|^2 \left \vert\theta_1(\bar{t})\right\vert^2
e^{-2\pi |\tau|\tilde{t}_1^2}\eta^{-2}.
\label{k02}
\eeqn
Then the only integration which is left is the integration with respect to the
toron field. From Eq.(\ref{fpt0}) it follows $(i,j =1,\ldots,r)$ :
\beqn
\det\left\Arrowvert S_t^{(0)}(x_i,y_j) \right\Arrowvert
=\left(\frac{\eta^3}{L_1}\right)^r e^{\frac{2\pi i}{|\tau|}\sum_{i=1}^r
(\zeta_i-\xi_i)\bar{t}}\det \left \|\frac{\theta_1(z_i-w_j+\bar{t})}
{\theta_1(z_i-w_j)\theta_1(\bar{t})} \right\|~.
\label{dett}
\eeqn
It can be proven that
\beqn
&&\det \left \|\frac{\theta_1(z_i-w_j+\bar{t})}{\theta_1(z_i-w_j)\theta_1
(\bar{t})} \right\|\cr
&&=(-1)^{\frac{r(r-1)}{2}}\frac{\prod_{i<j}^r \theta_1(z_i-z_j)
\theta_1(w_i-w_j)}{\theta_1(\bar{t})\prod_{i,j}^r\theta_1(z_i-w_j)}
\theta_1\left(\sum_{i=1}^r(z_i-w_i)+\bar{t} \right)~.
\label{Caudet}
\eeqn
This formula is a generalization to the torus of the Cauchy determinant
formula
\cite{Stone}
\beqn
\det \left \|\frac{1}{z_i-w_j} \right\|
=(-1)^{\frac{r(r-1)}{2}}\frac{\prod_{i<j}^r (z_i-z_j)
(w_i-w_j)}{\prod_{i,j}^r(z_i-w_j)}~.
\label{Cdf}
\eeqn
The proof is based on the examination of the zero and pole structure of the
l.h.s. of Eq.(\ref{Caudet}) using short distance behaviour given in l.h.s.
of Eq.(\ref{Cdf}). Then one may check that functions in both sides obey
the same periodicity conditions when $z_i\rightarrow z_i+1, w_j\rightarrow
w_j+1$ and $z_i\rightarrow z_i+\tau, w_j\rightarrow w_j+\tau$ (standard
elliptic function arguments).
\par
Now the integration with respect to the toron field can be done
with the help of the formula
\beqn
\int_0^1d\tilde{t}_1\int_0^1d\tilde{t}_2e^{4\pi \zeta\tilde{t}_1}
\theta_a(z+\bar{t})\theta_a(\bar{z}+t)e^{-2\pi|\tau|\tilde{t}_1^2} =
\frac{e^{\frac{2\pi \zeta^2}{|\tau|}}}{\sqrt{2|\tau|}}, ~~~a=1,3
\label{torint1}
\eeqn
and we get
\beqn
&&Z_t^{-1}\int_0^{T_1}dt_1\int_0^{T_2}dt_2
\left |\det \left \|S^{(0)}_t(x_i,y_j)\right \|
\right|^2 |\theta_1(\bar{t})|^2e^{-2\pi |\tau|\tilde{t}_1^2}\eta^{-2}\cr
&&=\left(\frac{\eta^3}{L_1}\right)^{2r}\frac{1}{\sqrt{2|\tau|}}
e^{-\frac{2\pi}{|\tau|}\left \{ \sum_{i<j}^r
\left [(\zeta_i-\zeta_j)^2+(\xi_i-\xi_j)^2\right]-\sum_{i=1}^r\sum_{j=1}^r
(\zeta_i-\xi_j)^2\right\}}\cr
&&\times \frac{\prod_{i<i'}^r|\theta_1(z_i-z_{i'})|^2
\prod_{j<j'}^r|\theta_1(w_j-w_{j'})|^2}{\prod_{i=1}^r\prod_{j=1}^r
|\theta_1(z_i-w_j)|^2}~.
\label{torint2}
\eeqn
If we insert this result into Eq.(\ref{k02}) and take into account
Eq.(\ref{G0prop}) together with Eqs (\ref{fc}) and (\ref{pft})
we will see that this is Eq.(\ref{genfor}) for the case when $r=s$.

2) $r-s=k>0$. Only a sector with the topological charge $k$
contributes and Eq.(\ref{genfor}) takes a form \beqn &&\la
\prod_{i=1}^r\bar{\psi}(x_i)P_+\psi(x_i)
\prod_{j=1}^s\bar{\psi}(y_j)P_-\psi(y_j)\ra
=\left(\la\bar{\psi}(x)P_+\psi(x)\ra \right)^N \cr &&\times
e^{-4\pi\sum_{i<i'}^r
\overline{G}_m(x_i-x_{i'})-4\pi\sum_{j<j'}^s\overline{G}_m(y_j-y_{j'})
+4\pi\sum_{i=1}^r\sum_{j=1}^s\overline{G}_m(x_i-y_j)}.
\label{gforrs} \eeqn From the general formula Eq.(\ref{gfor}) for
$s\geq 1$ it follows \beqn &&\la
\prod_{i=1}^r\bar{\psi}(x_i)P_+\psi(x_i)
\prod_{j=1}^s\bar{\psi}(y_j)P_-\psi(y_j)\ra \cr
&&=Z^{-1}L_1^k\int_{{\cal A}_k}{\cal D}Ae^{-S[A]+\frac{1}{2}
\Gamma^{(k)}_{\rm reg}[A]} \left \vert \det\left\Arrowvert
(\hat{\chi}, S^{(k)})\right\Arrowvert \right\vert^2\left(\det{\cal
N}_A^{(k)}\right)^{-1}~, \label{rs1} \eeqn where the $r\times r$
matrix $\left \Arrowvert (\hat{\chi}, S^{(k)})\right\Arrowvert$
reads as \beqn \left \Arrowvert (\hat{\chi},
S^{(k)})\right\Arrowvert =\left(\matrix{
\hat{\chi}_1^{(1)}(x_1)&\ldots&\hat{\chi}_1^{(k)}(x_1)
&S^{(k)}(x_1,y_1;A)&\ldots &S^{(k)}(x_1,y_s;A)\cr \vdots &\ddots
&\vdots &\vdots &\ddots &\vdots\cr \hat{\chi}_1^{(1)}(x_r)&\ldots
&\hat{\chi}_1^{(k)}(x_r) &S^{(k)}(x_r,y_1;A)&\ldots
&S^{(k)}(x_r,y_s;A) \cr}\right)~. \label{chiSmat} \eeqn Now we may
use the relation between the zero modes $\hat{\chi}(x)$ and
$\chi(x)$, Eq.({\ref{fermprop}), and do the path integration over
the $b(x)$ field with the result \beqn &&\la
\prod_{i=1}^r\bar{\psi}(x_i)P_+\psi(x_i)
\prod_{j=1}^s\bar{\psi}(y_j)P_-\psi(y_j)\ra
=Z^{-1}_te^{N\left(2e^2G(0)-\frac{2\pi ^2}{e^2L_1L_2} \right)}\cr
&&\times e^{-4\pi\sum_{i<i'}^r
\overline{G}_m(x_i-x_{i'})-4\pi\sum_{j<j'}^s\overline{G}_m(y_j-y_{j'})
+4\pi\sum_{i=1}^r\sum_{j=1}^s\overline{G}_m(x_i-y_j)}\cr &&\times
e^{4\pi\sum_{i<i'}^r G_0(x_i-x_{i'})+4\pi\sum_{j<j'}^s
G_0(y_j-y_{j'}) -4\pi\sum_{i=1}^r\sum_{j=1}^s G_0(x_i-y_j)}\cr
&&\times \int_0^{T_1}dt_1\int_0^{T_2}dt_2 \left\vert
\det\left\Arrowvert (\chi, S_t^{(k)})\right\Arrowvert
\right\vert^2 ~. \label{rs2} \eeqn From Eq.(\ref{fptk}) it follows
that \beqn &&\left\vert \det\left\Arrowvert (\chi,
S_t^{(k)})\right\Arrowvert \right\vert^2 =\left(\frac{2k}{|\tau|}
\right)^{k/2}\frac{\eta^{6s}}{L_1^{2k+2s}} e^{4\pi
\tilde{t}_1\left(\sum_{i=1}^r\zeta_i-\sum_{j=1}^s\xi_j\right)}\cr
&&\times e^{-\frac{2\pi k}{|\tau|}\left(\sum_{i=1}^r\zeta_i^2
-\sum_{j=1}^s\xi_j^2\right) - 2\pi |\tau|\tilde{t}_1^2}\vert\det
\Arrowvert M_{i,j}\Arrowvert \vert^2~, \label{detchiS} \eeqn where
$r\times r$ matrix $\Arrowvert M_{ij} \Arrowvert$ is defined such
that for $1\leq j \leq k$ \beqn M_{ij}= T_k^{(j)}(z_i+\bar{t}/k)
\label{matM1}
\eeqn and for $k+1\leq j \leq r$ \beqn
M_{ij}=\frac{\theta_1(z_i-w_{j-k}+\bar{t})q^{(k)}(z_i)}
{\theta_1(z_i-w_{j-k})\theta_1(\bar{t})q^{(k)}(w_{j-k})}~.
\label{matM2} \eeqn Now for the determinant  of the matrix $M$ it
can be proven the following expression \beqn \det\Arrowvert
M_{i,j}\Arrowvert = C_k \frac{\prod_{i<i'}^r
\theta_1(z_i-z_{i'})\prod_{j<j'}^s
\theta_1(w_j-w_{j'})}{\prod_{i=1}^r\prod_{j=1}^s\theta_1(z_i-w_j)}
\theta_a\left(\sum_{i=1}^r z_i-\sum_{j=1}^s w_j+\bar{t} \right)~~,
\label{detM} \eeqn where $a=1(3)$ if $k$ is even(odd). (If $s=1$
one should take 1 instead of the product $\prod_{j<j'}$in the
numerator.) For the cases $r=2, s=1 $ (considered in \cite{SSZ})
and $r=3,s=1 $ we calculated the determinant in the l.h.s.
explicitely using Eqs (\ref{zeromode}), (\ref{fpt0}) and
(\ref{fptk}) and checked the formula Eq.(\ref{detM}).
\par
The proof of this formula for arbitrary values of $r$ and $s$ is again based
on the comparison of the analytic structures and periodicity properties of
its both sides. The constant $C_k$ can not be fixed by
this consideration and in order to find it one should do some additional
analysis.
\par
Now with the help of Eq.(\ref{torint1}) we can do the integration
with respect to the toron field and obtain \beqn
&&\int_0^1d\tilde{t}_1\int_0^1d\tilde{t}_2 \left |\det \left
\|(\chi,S^{(k)}_t)\right \| \right|^2=
|C_k|^2\eta^{6s}\left(\frac{2k}{|\tau|}\right)^{k/2}
\frac{1}{\sqrt{2|\tau|} L_1^{2k+2s}}\cr &&\times \exp \left
\{-\frac{2\pi}{|\tau|}\left [\sum_{i<i'}^r (\zeta_i-\zeta_{i'})^2
+\sum_{j<j'}^s(\xi_j-\xi_{j'})^2-\sum_{i=1}^r
\sum_{j=1}^s(\zeta_i-\xi_j)^2\right] \right\}\cr &&\times
\frac{\prod_{i<i'}^r|\theta_1(z_i-z_{i'})|^2
\prod_{j<j'}^s|\theta_1(w_j-w_{j'})|^2}{\prod_{i=1}^r\prod_{j=1}^s
|\theta_1(z_i-w_j)|^2}~. \label{intdetchiS} \eeqn As we see for
our aim it is sufficient to know only $|C_k|^2$. In order to find
it we may use the properties which the objects entering
Eq.(\ref{intdetchiS}) demonstrate under the modular transformation
considered in Section 3. The l.h.s. of Eq.(\ref{intdetchiS}) is
invariant under this transformation so its r.h.s. should be
invariant as well. With the help of Eq.(\ref{G0prop}) we find that
it will really be the case if \beqn |C_k|^2 =\eta^{-(k-1)(k-2)}~~.
\label{constC} \eeqn Using this expression together with
Eqs.(\ref{intdetchiS}), (\ref{fc}) and (\ref{pft}) in
Eq.(\ref{rs2}) we will get the desired result (Eq.(\ref{gforrs})).
\par
The case when $s=0$ can be considered similarly. Now instead of the matrix
Eq.(\ref{chiSmat}) we will have a matrix of the zero modes
$\left \Arrowvert\chi_1^{(j)}(x_i)\right \Arrowvert $ only.
To obtain the result in this case we may use the formula
\beqn
\det\left\Arrowvert T_k^{(j)}(z_i+\bar{t}/k)\right\Arrowvert
=C_k \prod_{i<j}^k\theta_1(z_i-z_j)\theta_a\left(\sum_{i=1}^kz_i+\bar{t}\right)
\eeqn
where $a=1(3)$ if $k$ is even (odd). This formula can be proven
by the same method as the formula Eq.(\ref{detM}).
\section{Conclusions}
Many interesting features of the SM on a torus are related to the
fact that on the torus one can separate in a simple manner the
zero modes from the other degrees of freedom. They need a special
treatment in the quantum theory and contribute to correlation
functions of the fermion fields. The role of the zero modes in the
chiral symmetry breaking by an anomaly and in the occurence of
clustering becomes particularly transparent.
\par
The dynamics of the toron field is determined by the  action
$\Gamma^{(0)}[t]$, which is induced by the effect of this field on the
fermions. It controls infrared singularities. The averiging with respect
to the toron field assures a translation invariant distribution of the
symmetry breaking zero modes in the topologically non-trivial sectors.
\par
The knowledge of the exact expressions of the N-point correlators is
necessary for finding of the finite temperature spectral functions and
offer important information related to the symmetry problems \cite{Hansson}.

\par
There are several interesting issues possible for the further investigation.
\par
One can extend our consideration to the case of the still exactly soluble
geometric ($N_f=2$) \cite{JA} and multiflavor ($N_f$ is arbitrary) massless
SM \cite{Gatt}, where in the spectrum in addition to one massive particle
there appear an iso-spin multiplet of massless particles. In this case the
factor $N_f$ which appears in the toron action will change the character of
the toron integration considerably and hence their dynamical role.
\par
The general formula Eq.(\ref{genfor}) which we obtained in the present work
is exremely useful for the consideration of the two-dimensional QED with one
flavor massive fermions (massive SM). This model is not exactly soluble but
one can do the perturbation expansion in the fermion mass following the
approach developed in \cite{massSM}.
\par
The perturbation theory in fermion masses cannot
be employed in the $N_f\geq2$ case as physical quantities are not analytic in
fermion mass at $T=0$ \cite{Coleman76},\cite{HH}. In this case it
can be applied only in the high temperature regime. In papers \cite{HH}
$QED_2$ with massive fermions on a circle have been investigated by the method of abelian
bosonization. We believe that new results in $QED_2$ with massive
N-flavor fermions will help to understand how the effect of quark masses
modify the vacuum structure, meson masses, mixing and the pattern of chiral
symmetry breaking.
\par
Another interesting problem is to consider SM on a torus at finite
density \cite{finden} and find out how the chemical potencial will enter
into our general formula Eq.(\ref{genfor}).
\par
A detailed discussion of the different limits $L_1,L_2 \rightarrow \infty$
is still to be done. For an useful discussion related to this problem see
\cite{Smilga1}.
\par
Although the present calculations have been done for a simple two-dimensional
abelian model we hope that they further our intuition needed to understand
non-perturbative physics of realistic theory such as QCD \cite{SMQCD}.
They could be useful for the comparison with the results obtained by the
authors who do the lattice simulations of the SM \cite{latSM}.
\par
Recently a systematic comparison between SM on a torus in the present
Euclidean (path integral) approach and SM on a circle in a Hamiltonian
(canonical) approach \cite{Man}, \cite{IM} has been fulfiled
\cite{JAinprep}.
It is worthwile to mention that the general formula Eq.(\ref{genfor})
can also be obtained in the second approach.\\ \\

{\bf Acknowledgments}\\

We would like to thank H.Joos for fruitful collaboration, reading
of the manuscript and helpful suggestions and A.Wipf for useful discussions.

\end{document}